\documentclass[aps,pra,superscriptaddress,notitlepage,twocolumn]{revtex4-2}

\usepackage{mathrsfs}
\usepackage{amsfonts}
\usepackage{amssymb}
\usepackage{amsmath}
\usepackage{bm}
\usepackage{graphicx}
\usepackage{sidecap}
\usepackage{color}
\usepackage[colorlinks=true,citecolor=blue,linkcolor=cyan]{hyperref}
\usepackage{subeqnarray}
\usepackage{verbatim}
\usepackage{ulem}
\usepackage{cases}
\usepackage{cancel}
\usepackage{framed}
\usepackage{tikz}
\usepackage{multirow}

\newcommand{\ket}[1]{\vert #1 \rangle}
\newcommand{\bra}[1]{\langle #1 \vert}
\newcommand{\ketbra}[2]{\vert #1 \rangle \langle #2 \vert}
\newcommand{\braket}[2]{\langle #1 \vert #2 \rangle}

\begin{document}
\title{Accelerating quantum adiabatic evolution with $\pi$-pulse sequences}
\author{Tonghao Xing}
\affiliation{Beijing Academy of Quantum Information Sciences, Beijing 100193, China}
\affiliation{State Key Laboratory of Low-Dimensional Quantum Physics and Department of Physics, Tsinghua University, Beijing 100084, China}
\author{Jiang Zhang}
\email{zhangjiang@baqis.ac.cn}
\affiliation{Beijing Academy of Quantum Information Sciences, Beijing 100193, China}
\date{\today}
\author{Guilu Long}
\email{gllong@mail.tsinghua.edu.cn}
\affiliation{Beijing Academy of Quantum Information Sciences, Beijing 100193, China}
\affiliation{State Key Laboratory of Low-Dimensional Quantum Physics and Department of Physics, Tsinghua University, Beijing 100084, China}
\affiliation{Frontier Science Center for Quantum Information, Beijing 100084, China}
\affiliation{Beijing National Research Center for Information Science and Technology, Beijing 100084, China}

\begin{abstract}
In quantum information processing, the development of fast and robust control schemes remains a central challenge. Although quantum adiabatic evolution is inherently robust against control errors, it typically demands long evolution times. In this work, we propose to achieve rapid adiabatic evolution, in which nonadiabatic transitions induced by fast changes in the system Hamiltonian are mitigated by flipping the nonadiabatic transition matrix using $\pi$ pulses. This enables a faster realization of adiabatic evolution while preserving its robustness. 
We demonstrate the effectiveness of our scheme in both two-level and three-level systems. Numerical simulations show that, for the same evolution duration, our scheme achieves higher fidelity and significantly suppresses nonadiabatic transitions compared to the traditional STIRAP protocol.
\end{abstract}

\maketitle
\section{introduction}
Quantum adiabatic evolution \cite{messiah2014quantum} is of fundamental interest in quantum physics, underpinning phenomena such as Abelian and non-Abelian adiabatic geometric phases \cite{berry1984quantal,wilczek1984appearance,zhang2023geometric} and Landau-Zener transitions \cite{landau1932theorie,zener1932non}.
Due to its intrinsic robustness against control errors, quantum adiabatic control has also found a wide range of applications in quantum technologies, including state engineering \cite{vitanov2001laser,vitanov2017stimulated,li2025topological}, quantum simulation \cite{aspuru2005simulated,kim2010quantum}, and quantum computation \cite{farhi2000quantum,farhi2001quantum,barends2016digitized,albash2018adiabatic}.

According to the adiabatic theorem \cite{messiah2014quantum}, a quantum process remains adiabatic, i.e., a state initially prepared in an eigenstate of the Hamiltonian stays close to the corresponding instantaneous eigenstate, if the system Hamiltonian varies slowly enough to satisfy the so-called adiabatic condition.
In its traditional form, this condition imposes a speed limit on adiabatic evolution, requiring the rate of 
change of the Hamiltonian to be much smaller than the energy gaps between eigenstates.
However, this condition has been shown to be neither sufficient nor necessary \cite{marzlin2004inconsistency,sarandy2004consistency,tong2005quantitative}, sparking significant debate over its validity \cite{amin2009consistency,tong2010quantitative} and prompting the development of several new adiabatic conditions \cite{tong2007sufficiency,wei2007quantum,comparat2009general,lidar2009adiabatic,boixo2010necessary,bachmann2017adiabatic,dodin2021generalized}.
Despite these advances, practical adiabatic control remains limited by the need for slow evolution.

Shortcuts to adiabaticity (STA) offers a promising solution by employing an auxiliary Hamiltonian to suppress nonadiabatic transitions induced by fast driving \cite{demirplak2003adiabatic,berry2009transitionless,chen2010shortcut,del2013shortcuts,zhang2015fast,santos2017generalized,guery2019shortcuts,del2019focus,abah2020quantum,chen2021shortcuts}.
To date, this approach has been experimentally demonstrated in various quantum systems, including ultracold atoms \cite{schaff2010fast,bason2012high,deng2018superadiabatic}, NV centers \cite{zhang2013experimental,kolbl2019initialization}, trapped ions \cite{an2016shortcuts}, and superconducting qubits \cite{vepsalainen2019superadiabatic}.
However, the auxiliary Hamiltonian, which corresponds to the adiabatic gauge potential \cite{kolodrubetz2017geometry,takahashi2024shortcuts}, often introduces additional transition terms that may be inaccessible in certain systems, thereby complicating experimental implementation \cite{ibanez2012multiple,martinez2014shortcuts,li2016shortcut,baksic2016speeding,du2016experimental,zhou2017accelerated}.
Moreover, introducing this auxiliary Hamiltonian drives the system away from the eigenstates of the original Hamiltonian, compromising the robustness of the adiabatic process.

To combine robustness and speed, it is desirable to maintain adiabatic evolution while accelerating the process through additional control techniques.
Recently, a $\pi$-pulse scheme based on a necessary and sufficient adiabatic condition is proposed for the two-level system \cite{wang2016necessary}.
Notably, this scheme enables adiabatic evolution even in the presence of vanishing energy gap by jumping across these zero-gap points.
This pioneering method has been experimentally verified \cite{xu2019breaking,zheng2022accelerated,gong2023accelerated} and applied to quantum sensing \cite{zeng2024wide}, electron-nuclear resonances enhancement \cite{xu2024enhancing}, state preparation \cite{chen2024fast}, and quantum computation \cite{zhang2025speeding}. However, it remains unclear whether this scheme can be generalized to a general form.

\begin{figure}[tbh]
	\centering
	\includegraphics[width=0.48\textwidth]{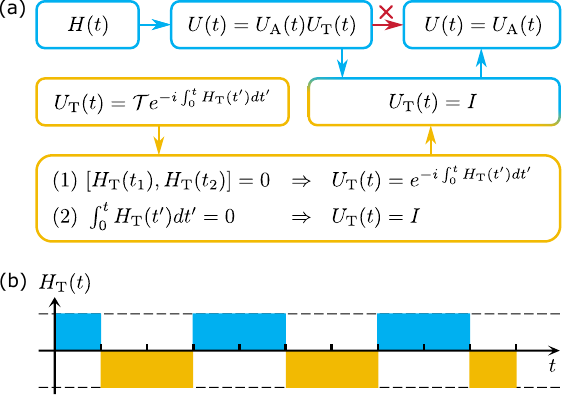}
	\caption{Schematic illustration of the method used to identify conditions for accelerating adiabatic evolution. (a) The system’s nonadiabatic transition operator $U_\text{T}(t)$ is generated by a Hermitian operator $H_\text{T}(t)$, which is determined by the time dependence of the system Hamiltonian $H(t)$. By identifying the conditions under which $H_\text{T}(t)$ leads to $U_\text{T}(t) = I$, one can guarantee that the evolution remains adiabatic. (b) A sequence of $\pi$ pulses is applied to repeatedly invert the sign of $H_\text{T}(t)$, effectively canceling its time-integrated contributions. As a result, $U_\text{T}(t)$ approaches the identity operator, suppressing nonadiabatic transitions.}\label{fig:schematic}
\end{figure}

In this work, we begin with a general evolution operator $U(t)$ governed by a system Hamiltonian $H(t)$. By decomposing $U(t)$ into a product of an adiabatic evolution operator $U_\text{A}(t)$ and a nonadiabatic transition operator $U_\text{T}(t)$, adiabaticity can be enforced by ensuring $U_\text{T}(t)$ approaches the identity. 
Using a two-step protocol (shown in Fig.~\ref{fig:schematic}(a)), we derive the conditions under which the system Hamiltonian guarantees such evolution. 
These conditions reveal a general scheme for accelerating adiabatic evolution, in which nonadiabatic transitions are compensated by employing piecewise constant nonadiabatic transition Hamiltonian $H_\text{T}(t)$, flipped using $\pi$ pulses that accumulate phase differences (shown in Fig.~\ref{fig:schematic}(b)).   
Our scheme is applicable to multi-level systems and allows evolution paths beyond geodesics.
We explicitly demonstrate this approach in two- and three-level systems. 
Specifically, we compare the performance of our scheme with the widely used stimulated Raman adiabatic passage (STIRAP) for state transfer. 
Numerical simulations show that our scheme yields higher target-state population than STIRAP, thus offering a novel and efficient strategy for realizing adiabatic evolution.

\section{quantum energy-level transition operator}
Consider an $N$-dimensional quantum system governed by a non-degenerate, time-dependent Hamiltonian,
\begin{equation}
	H(t)=\sum_n E_n(t)\ket{n(t)}\bra{n(t)}.
\end{equation}
The system's dynamics at time $t$ are described by the time evolution operator $U(t)$, which satisfies the following equation ($\hbar=1$)
\begin{equation}
	i\dot{U}(t)=H(t)U(t).
\end{equation}

According to the adiabatic theorem, if the system is initially in an eigenstate of the Hamiltonian, e.g., $\ket{n(0)}$, it will remain in the corresponding instantaneous eigenstate  $\ket{n(t)}$ at time $t$, provided that $H(t)$ varies slowly enough to satisfy the adiabatic condition.
However, when $H(t)$ changes rapidly, transitions between energy levels may occur, and the system may no longer stay in the state $\ket{n(t)}$.

To analyze energy-level transitions induced by $H(t)$, we decompose the evolution operator $U(t)$ into the product of an adiabatic unitary operator and a unitary operator describing transitions, so that the effects of nonadiabatic transitions can be isolated. Explicitly, we write $U(t)=U_\text{A}(t)U_\text{T}(t)$, where $U_\text{A}(t)$ and $U_\text{T}(t)$ denote the adiabatic and transition unitaries, respectively.

The adiabatic part can be further expressed as $U_\text{A}(t) = U_d(t)U_g(t)$, where
\begin{equation}
	U_d(t) = \sum_n e^{-i\int_0^t E_n(t')\,\mathrm{d}t'} \ket{n(t)}\bra{n(0)}
\end{equation}
is the dynamical phase contribution, and
\begin{equation}
	U_g(t) = \sum_n e^{-\int_0^t \braket{n(t')}{\dot{n}(t')} \,\mathrm{d}t'} \ket{n(0)}\bra{n(0)}
\end{equation}
represents the geometric phase contribution.

Accordingly, the nonadiabatic transition operator $U_\text{T}(t)$ is given by 
\begin{equation}\label{eq:UTt}
	U_\text{T}(t)=\mathcal{T}e^{-i\int_0^tH_\text{T}(t')\mathrm{d}t'},
\end{equation}
where $\mathcal{T}$ denotes the time-ordering operator. The effective transition Hamiltonian $H_\text{T}(t)$ reads
\begin{align}\label{eq:HTt}
	H_\text{T}(t)=-\sum_{n\neq m}&e^{i\int_0^{t}E_n(t')-E_m(t')\mathrm{d}t'} \nonumber \\
	\cdot &e^{\int_0^{t}\braket{n(t')}{\dot{n}(t')}-\braket{m(t')}{\dot{m}(t')}\mathrm{d}t'} \nonumber \\
	\cdot &i\braket{n(t)}{\dot{m}(t)}\ketbra{n(0)}{m(0)}.
\end{align}
If $H_\text{T}(t)$ satisfies appropriate conditions such that $U_\text{T}(t)=I$, i.e., the identity operator, then the system undergoes no transitions between the instantaneous eigenstates. In this case, the overall evolution can be regarded as adiabatic.

\section{Rapid Adiabatic Evolution Conditions}

As shown in Eq.~(\ref{eq:HTt}), the nonadiabatic transition Hamiltonian $H_\text{T}(t)$ is determined by the system Hamiltonian $H(t)$. Consequently, from Eq.~(\ref{eq:UTt}), if one can design a Hamiltonian $H(t)$ such that $U_\text{T}(t) = I$, then no energy-level transitions occur during the evolution. In this case, the evolution can be considered adiabatic.

However, for a general time-dependent Hamiltonian $H(t)$, obtaining $U_\text{T}(t)$ is typically intractable. To address this challenge, we adopt a two-step protocol to ensure that $U_{\mathrm{T}}(t) = I$.  
In the first step, we show that when $H(t)$ satisfies certain conditions, the Hamiltonians $H_\text{T}(t)$ at different times commute with each other. In such cases, the evolution operator $U_\text{T}(t)$ becomes exactly solvable. In the second step, we identify the conditions under which $U_\text{T}(t)$ becomes the identity operator.

To simplify the expressions for $H_\text{T}(t)$ and $U_\text{T}(t)$, we define the following quantities:
\begin{align}\label{eq:egphi}
	E_{n,m}(t) &= E_n(t) - E_m(t), \nonumber \\
	g_{n,m}(t) &= i\braket{n(t)}{\dot{m}(t)}, \nonumber \\
	\phi_{n,m}(t) &= \int_0^t \left[ E_{n,m}(t') - \left(g_{n,n}(t') - g_{m,m}(t')\right) \right] \mathrm{d}t'.
\end{align}
With these definitions, the Hamiltonian $H(t)$ becomes
\begin{equation}
	H_\text{T}(t) = -\sum_{n \neq m} e^{i\phi_{n,m}(t)} g_{n,m}(t) \ketbra{n(0)}{m(0)}.
\end{equation}

Following these formulations, the first step is to determine the conditions under which
\begin{equation}
[H_\text{T}(t_i), H_\text{T}(t_j)] = 0,
\end{equation}
for arbitrary distinct times $t_i$ and $t_j$.
The second step is to identify the conditions under which
\begin{equation}\label{eq:int}
	\int_0^t e^{i\phi_{n,m}(t')} g_{n,m}(t') \,\mathrm{d}t' = 0, \quad \forall\, n \neq m.
\end{equation}
Any Hamiltonian $H(t)$ that satisfies the conditions from both steps yields an adiabatic evolution.

\subsection{The First Step: Conditions for Commutativity}

We begin by investigating the conditions under which $H_\text{T}(t)$ commutes with itself at different times. A straightforward calculation shows that the commutator $[H_\text{T}(t_1), H_\text{T}(t_2)] = 0$ imposes the following constraint for each combination of indices $\{n, m, p\}$ and times $\{t_1, t_2\}$:
\begin{align}
	&e^{i[\phi_{n,m}(t_1)+\phi_{m,p}(t_2)]} g_{n,m}(t_1) g_{m,p}(t_2) \nonumber \\
	-&e^{i[\phi_{n,m}(t_2)+\phi_{m,p}(t_1)]} g_{n,m}(t_2) g_{m,p}(t_1) = 0,
\end{align}
where $n \neq m$ and $m \neq p$, although $n$ is allowed to equal $p$. For example, in a two-level system, one can consider the case $n=1$, $m=2$, and $p=1$.

To satisfy this equation, it is necessary and sufficient that both its real and imaginary parts vanish. This leads to the following system of equations:
\begin{equation}\label{eq:sin}
	\left\{
	\begin{aligned}
		&\cos[\phi_{n,m}(t_1)+\phi_{m,p}(t_2)]\, g_{n,m}(t_1) g_{m,p}(t_2) \\
		&= \cos[\phi_{n,m}(t_2)+\phi_{m,p}(t_1)]\, g_{n,m}(t_2) g_{m,p}(t_1), \\
		&\sin[\phi_{n,m}(t_1)+\phi_{m,p}(t_2)]\, g_{n,m}(t_1) g_{m,p}(t_2) \\
		&= \sin[\phi_{n,m}(t_2)+\phi_{m,p}(t_1)]\, g_{n,m}(t_2) g_{m,p}(t_1).
	\end{aligned}
	\right.
\end{equation}
This system provides two constraints involving eight variables, thereby allowing infinitely many solutions.

It is important to note that while some of these solutions are mathematically valid, they may be physically trivial or uninteresting. For example, setting $g_{n,m}(t) = 0$ for all $n \neq m$ and for all $t$ trivially satisfies Eq.~(\ref{eq:sin}). However, this condition implies that the eigenstates of $H(t)$ remain stationary, leading to no evolution in the system.
This observation illustrates that to obtain physically meaningful solutions, the constraints on $g_{n,m}(t)$ must be relaxed. One approach is to impose conditions on $\phi_{n,m}(t)$ and the angle difference
\begin{equation}
	\gamma_{n,m}^{m,p}(t_1,t_2) = [\phi_{n,m}(t_1)+\phi_{m,p}(t_2)] - [\phi_{n,m}(t_2)+\phi_{m,p}(t_1)],
\end{equation}
which is determined solely by the phases.

From the definitions in Eq.~(\ref{eq:egphi}), one can verify the following properties of $\phi_{n,m}(t)$:
\begin{align}\label{eq:phinm}
	&\text{(i)} \quad \phi_{n,m}(t) + \phi_{m,p}(t) = \phi_{n,p}(t), \nonumber \\
	&\text{(ii)} \quad \phi_{n,m}(t_2) - \phi_{n,m}(t_1) = \phi_{n,m}(t_1, t_2),
\end{align}
where $\phi_{n,m}(t_1, t_2)$ denotes the integral of the integrand in $\phi_{n,m}(t)$ from $t_1$ to $t_2$.
Here, we define $\phi_{n,m}(t) = a t (n - m) \pi$, where $a$ is a constant. This definition satisfies both properties in Eq.~(\ref{eq:phinm}). Consequently, the angle difference becomes
\begin{equation}
	\gamma_{n,m}^{m,p}(t_1, t_2) = a(n - 2m + p)(t_2 - t_1)\pi.
\end{equation}

To relax the constraints on $g_{n,m}(t)$, we require that $\gamma_{n,m}^{m,p}(t_1,t_2)=0$ or $\pi$ (modulo $2\pi$), which implies that $g_{n,m}(t_1)g_{m,p}(t_2)=\pm g_{n,m}(t_2)g_{m,p}(t_1)$.
This condition can be satisfied by defining $\phi_{n,m}(t)$ as a piecewise function taking discrete values:
\begin{equation}
	\phi_{n,m}(t)=\left\{\begin{aligned}
		&0, \quad &&t\in[0,t_1); \\
		&a(n-m)\pi, \quad &&t\in[t_1,t_2); \\
		&\cdots, \quad &&\cdots; \\
		&aK(n-m)\pi, \quad &&t\in[t_K,\tau],
	\end{aligned}
	\right.
\end{equation}
where $a$ is an integer, and the total evolution time $\tau$ is divided into $K+1$ segments.
For convenience, we write $\phi_{n,m}(t)=aj(n-m)\pi$ within each time interval $t\in[t_{j},t_{j+1})$, leading to:
\begin{equation}
	\gamma_{n,m}^{m,p}(t_j,t_l)=a(n-2m+p)(l-j)\pi.
\end{equation}

When $a$ is odd, $\gamma_{n,m}^{m,p}(t_j,t_l)$ depends on both the energy levels $\{n,m,p\}$ and the time interval. In this case, we define $g_{n,m}(t)$ as: 
\begin{align}
	g_{n,m}(t_j)=\left\{\begin{aligned}
		&g_{n,m}, \quad &&n-m \:\text{is odd}; \\
		&0, \quad &&n-m\: \text{is even}.
	\end{aligned}
	\right.
\end{align}
Alternatively, we may set $g_{n,m}(t)=0$ for odd $n-m$ and $g_{n,m}(t)=g_{n,m}$ for even $n-m$. In this scenario, we have removed the dependence from $\gamma_{n,m}^{m,p}(t_j,t_l)$, which simplifies the derivation of a system Hamiltonian. However, as we will show in the second step, the latter choice cannot guarantee that $U_{\text{T}}(t)=I$, and will therefore not be considered further.

If $a$ is even, then $\phi_{n,m}(t_j)=\gamma_{n,m}^{m,p}(t_j,t_l)=0$ (modulo $2\pi$). We will show in the second step that in this scenario $U_{\text{T}}(t)$ cannot equal the identity unless $g_{n,m}(t)=0$, which is physically meaningless. Thus, we conclude that $a$ must be odd for our purposes.

\begin{table}
	\centering
	\caption{Three distinct cases considered in the second step. In Case 1, $a$ is even. In Cases 2 and 3, $a$ is set to 1 for simplicity and $g_{n,m}$ is a non-zero constant.}\label{tab:3cases}
	\begin{tabular}{cccc}
		\hline\hline 
		 &$\phi_{n,m}(t)$ & \multicolumn{2}{c}{$g_{n,m}(t)$}  \\ \cline{3-4}
		 &                  & $(n-m)$ even& $(n-m)$ odd \\ \hline
		Case 1\;~&         0        & unspecified   & unspecified \\
		Case 2\;~&      $j(n-m)\pi$ & $g_{n,m}$   & 0  \\
		Case 3\;~&      $j(n-m)\pi$ & 0           & $g_{n,m}$ \\ \hline\hline
	\end{tabular}
\end{table}

In summary, three representative cases will be analyzed in the next step. Their parameter configurations are listed in Table~\ref{tab:3cases}, where $g_{n,m}$ denotes a non-zero constant. As we shall demonstrate, while the first two cases do not ensure that $U_{\text{T}}(t) = I$, the third case provides a promising path toward constructing a physically realizable control Hamiltonian for fast adiabatic evolution.

\subsection{The second step: Conditions for the integral to equal zero}
To ensure that $U_{\text{T}}(t)=I$, we require the integral $\int_0^t e^{i\phi_{n,m}(t')} g_{n,m}(t') \mathrm{d}t' = 0$. In the following, we examine each case presented in Table~\ref{tab:3cases} to determine whether this condition can be satisfied.

\textbf{Case 1}. When $a$ is even, we have $\phi_{n,m}(t)=0$, so that $e^{i\phi_{n,m}(t)}=1$. In  this case, the condition reduces to
\begin{equation}
	\int_0^t g_{n,m}(t') \mathrm{d}t' = 0,
\end{equation}
which implies $g_{n,m}(t)=0$. However, this solution lacks physical meaning and is not viable.

\textbf{Case 2}. Here, $\phi_{n,m}(t)=j(n-m)\pi$ for $t\in[t_j,t_{j+1})$. When $n-m$ is odd, $g_{n,m}(t)=0$; when $n-m$ is even, $g_{n,m}(t)=g_{n,m}$. The integral becomes  
\begin{align}
	&\int_0^t e^{i\phi_{n,m}(t')} g_{n,m}(t') \mathrm{d}t'=\sum_{j=0}^{K} e^{i\phi_{n,m}(t_j)} g_{n,m}(t_j)\Delta(j),
\end{align}
where $\Delta(j)=t_{j+1}-t_{j}$ and $\sum_{j}\Delta(j)=\tau$.
When $n-m$ is even,  we  have $g_{n,m}(t)=g_{n,m}$, and the integral becomes 
\begin{equation}
	\sum_{j=0}^{K} g_{n,m}\Delta(j)=g_{n,m}\tau\neq 0.
\end{equation}
Therefore, this case does not satisfy the condition $U_{\text{T}}(t)=I$.

\textbf{Case 3}. In this case, for $t\in[t_j,t_{j+1})$, we have $\phi_{n,m}(t)=j(n-m)\pi$, but $g_{n,m}(t)=g_{n,m}$ when $n-m$ is odd, and $g_{n,m}(t)=0$ when $n-m$ is even.
If $n-m$ is even, $\phi_{n,m}(t)=0$ and $g_{n,m}(t)=0$, so the integral is zero.
If $n-m$ is odd, $\phi_{n,m}(t)=0(\pi)$ when $j$ is even (odd) and $g_{n,m}(t)=g_{n,m}$.
The integral then becomes
\begin{align}\label{eq:integral}
	&\int_0^t e^{i\phi_{n,m}(t')} g_{n,m}(t') \mathrm{d}t'=\sum_{j=0}^{K} e^{i\phi_{n,m}(t_j)} g_{n,m}\Delta(j) \nonumber \\
	=&\sum_{j=0}^{K} (-1)^jg_{n,m}\Delta(j),
\end{align}
which indicates the condition $\sum_{j=0}^{K} (-1)^jg_{n,m}\Delta(j)=0$.

Combing the two steps, we obtain the following conditions for the system Hamiltonian:
\begin{equation}\label{eq:conditions}
	\begin{cases}
		\phi_{n,m}(t)=j(n-m)\pi, \,\, \text{for} \,\, t\in[t_j,t_{j+1});\\
		g_{n,m}(t)=\begin{cases}
			g_{n,m},  & n-m \:\text{is odd}; \\
			0,  & n-m\: \text{is even};
		\end{cases} \\
		\sum_{j=0}^{K} (-1)^jg_{n,m}\Delta(j)=0;
	\end{cases}
\end{equation}
in which $n\neq m$.
This is the main result of this work.

The remaining task is to induce a flip in $\phi_{n,m}(t)$ between $\pi$ and $0$ (modulo $2\pi$) at each switching time instant $t_j$. This can be accomplished by turning on the Hamiltonian $H(t_j)$ for short duration $\delta_j$ before each $t_j$ ($j\ge1$), such that $\int_{t_j-\delta_j}^{t_j}\phi_{n,m}(t)\mathrm{d}t=\pi$.
Note that during $\delta_j$, $H(t_j)$ is time-independent, and thus all $g_{n,m}$ vanish, ensuring no transition errors accumulate as per Eq.~(\ref{eq:integral}). The total evolution time is therefore given by 
\begin{equation}
	\tau_t=\tau+\sum_{j=1}^{K}\delta_j.
\end{equation}

To maintain $\phi_{n,m}(t)$ as a constant over a time interval $[t_j, t_{j+1})$, two scenarios must be considered.
In the first scenario, the condition $g_{n,n}(t)-g_{m,m}(t) = 0$ holds. Then $\phi_{n,m}(t)$ remains constant only if $E_{n,m}(t) = 0$, which implies that no Hamiltonian $H(t)$ is applied during this segment. In this case, the system Hamiltonian is only active at discrete points where $\phi_{n,m}(t)$ flips.
In the second scenario where $g_{n,n}(t)-g_{m,m}(t) \neq 0$, to keep $\phi_{n,m}(t)$ constant, it is necessary that $E_{n,m}(t) = g_{n,n}(t) - g_{m,m}(t)$. Thus, a control Hamiltonian must be continuously applied along the  evolution path to meet the adiabatic condition.
In the next section, we will illustrate how these conditions can be explicitly realized in two- and three-level systems.

Finally, it is important to emphasize that while the integral in Eq.~(\ref{eq:integral}) vanishes at time $t$, it may not be zero at all times smaller than $t$. This means that $U_{\text{T}}$ may deviate from the identity operator $I$ at intermediate times. The maximum deviation generally depends on the number of segments $K+1$ used in the evolution path. The larger the value of $K$, the smaller the maximum deviation. Therefore, to suppress nonadiabatic transitions to an acceptable level, an suitable value of $K$ must be chosen. Concrete examples will be provided in the following section.

\section{Applications to two- and three-level systems}

\subsection{Two-level system}
We now apply the general conditions derived in Eq.~(\ref{eq:conditions}) to a concrete two-level quantum system. This analysis demonstrates how appropriate design of the control trajectory and system Hamiltonian can suppress nonadiabatic transitions while ensuring the desired unitary evolution.

For a two-level system, the general time-dependent Hamiltonian can be written as
\begin{equation}
	H(t)=\frac{\Omega(t)}{2}\left(\cos\theta\sigma_z + \sin\theta\cos\varphi\sigma_x + \sin\theta\sin\varphi\sigma_y\right),
\end{equation}
where $\Omega(t)$ is the pulse envelope, and $\theta$ and $\varphi$ are time-dependent control parameters defining the direction of the effective field on the parameter sphere. The operators $\sigma_{x,y,z}$ are the standard Pauli matrices.

This Hamiltonian has instantaneous eigenvalues $E_1=\Omega(t)/2$ and $E_2=-\Omega(t)/2$. The corresponding orthonormal eigenstates are
\begin{align}
	\ket{1(t)} &= \cos\frac{\theta}{2}\ket{g} + e^{i\varphi}\sin\frac{\theta}{2}\ket{e}, \nonumber \\
	\ket{2(t)} &= \sin\frac{\theta}{2}\ket{g} - e^{i\varphi}\cos\frac{\theta}{2}\ket{e},
\end{align}
where $\ket{g}$ and $\ket{e}$ denote the bare ground and excited states, respectively.

The energy gap is $E_{1,2}(t) = E_1 - E_2 = \Omega(t)$. The nonadiabatic couplings are calculated as
\begin{align}
	g_{1,1}(t) &= \frac{1}{2}(\cos\theta - 1)\dot{\varphi}, \quad
	g_{2,2}(t) = -\frac{1}{2}(1 + \cos\theta)\dot{\varphi}, \nonumber \\
	g_{1,2}(t) &= g_{2,1}^* = \frac{1}{2}\left(\dot{\varphi}\sin\theta + i\dot{\theta}\right).
\end{align}
The corresponding $\phi_{1,2}(t)$ is given by
\begin{equation}\label{eq:phi12}
	\phi_{1,2}(t) = -\phi_{2,1}(t) = \int_0^t \left[\Omega(t') - \cos\theta(t')\dot{\varphi}(t')\right] \mathrm{d}t'.
\end{equation}

To satisfy the constraints in Eq.~(\ref{eq:conditions}), we analyze the case where the off-diagonal coupling $g_{1,2}(t)$ is a constant complex number, denoted as $C = c_\text{r} + i c_\text{i}$. This leads to the following relations:
\begin{equation}
	\dot{\theta} = 2c_\text{i}, \qquad \dot{\varphi} = \frac{2c_\text{r}}{\sin\theta}.
\end{equation}
Many common evolution paths satisfy these relations naturally. Below, we list three representative cases:
\begin{itemize}
	\item[(i)] \textbf{Fixed meridian:} $\theta(t) = \theta(0) + c_\text{i}t$, $\varphi(t) = \varphi(0)$. This corresponds to evolution along a longitudinal circle (meridian) on the Bloch sphere.
	
	\item[(ii)] \textbf{Fixed parallel:} $\theta(t) = \theta(0)$, $\varphi(t) = \varphi(0) + \frac{2c_\text{r}t}{\sin\theta}$. This describes evolution along a circle of latitude (parallel).
	
	\item[(iii)] \textbf{General smooth path:} $\theta(t) = \theta(0) + c_\text{i}t$, $\varphi(t) = \varphi(0) + \int_0^t \frac{2c_\text{r}}{\sin\theta(t')} \mathrm{d}t'$, which allows for simultaneous smooth variations in both parameters.
\end{itemize}

Once a path is given, i.e., $\theta$ and $\varphi$ are specified, the pulse envelop $\Omega(t)$ can be determined using Eqs.~(\ref{eq:conditions}) and (\ref{eq:phi12}).
Specifically, we require that $\Omega(t)=\cos\theta\dot{\varphi}$ during the evolution, except at the time intervals $\delta_j$.
Additionally, $\phi_{1,2}(t)$ must accumulate a phase factor of $\pi$ during each $\delta_j$.
This design ensures the off-diagonal contribution from $g_{1,2}(t)$ integrates to zero over the total evolution, effectively suppressing unwanted transitions.

The so-called “jump protocol” proposed in Ref.~\cite{wang2016necessary} corresponds to a special case of scenario (ii) with $\theta = \pi/2$, i.e., evolution confined to the equator of the Bloch sphere. In this case, $\Omega(t)=0$ along the trajectory, and all the phase accumulation is concentrated at discrete time points via instantaneous Hamiltonian ``kicks''. Our framework generalizes this concept by enabling continuous and smooth evolution while maintaining the same transition suppression principles.

\begin{figure}[tbh]
	\centering
	\includegraphics[width=0.48\textwidth]{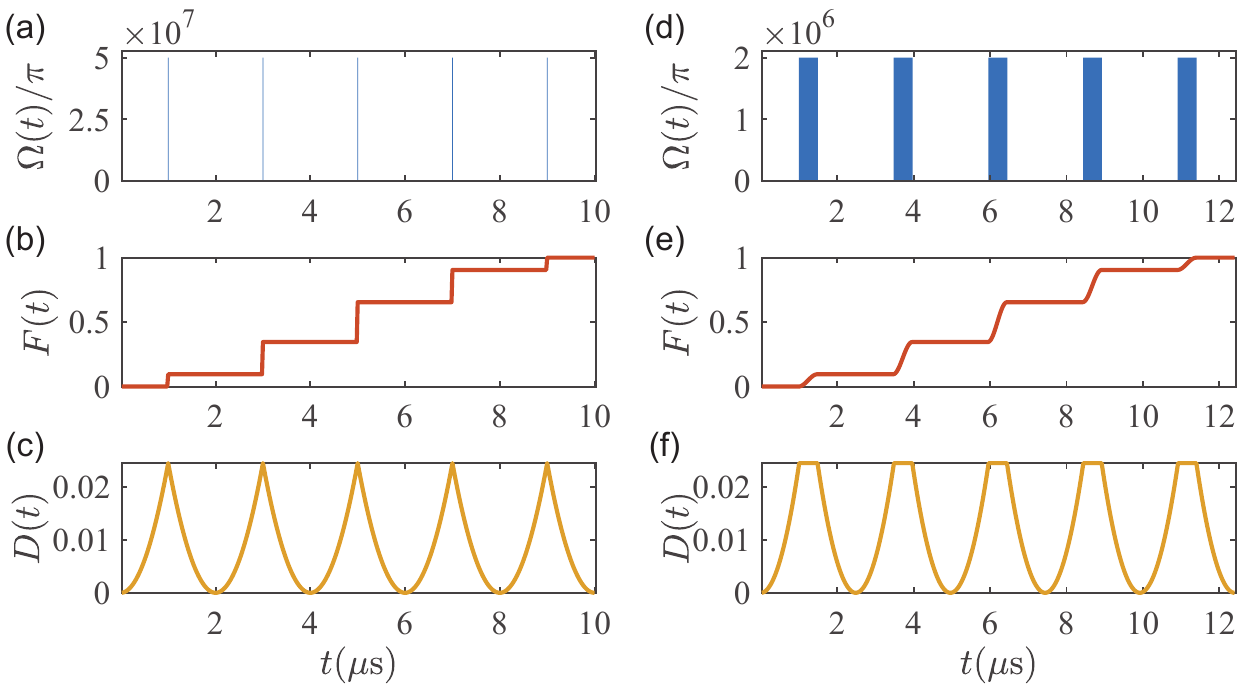}
	\caption{Accelerated adiabatic evolution along the path in Case (i). (a–c): ideal scenario with exact pulse strength, showing (a) the pulse envelope $\Omega(t)$, (b) the fidelity of the evolved state with respect to the instantaneous eigenstate, and (c) the adiabatic deviation function. (d–f): corresponding results under practical pulse shaping constraints.}\label{fig:case1}
\end{figure}

\subsection{Numerical simulations for two-level system}

To demonstrate the effectiveness of our scheme in suppressing nonadiabatic transitions, we perform numerical simulations of the two-level system dynamics. We evaluate the system performance using two quantitative metrics: the state fidelity $F(t)$ and the adiabatic path deviation function $D(t)$, both of which capture different aspects of the evolution's adiabaticity.

The state fidelity is defined as 
\begin{equation}
	F(t)=|\bra{\psi_i}U(t)^{-1}\ket{\psi_f}|^2,
\end{equation}
where $\ket{\psi_i}$ is the initial state, $\ket{\psi_f}$ is the target final state under ideal evolution, and $U(t) = \mathcal{T} \exp\left(-i\int_0^t H(t')\,dt'\right)$ is the time evolution operator. This quantity captures the overlap between the actual final state and the desired target state.

The adiabatic path deviation function is defined as 
\begin{equation}
	D(t) = 1-|\bra{\psi_i}U(t)^{-1}\ket{\psi_a(t)}|^2,
\end{equation}
where $\ket{\psi_a(t)} = U_A(t)\ket{\psi_i}$, and $U_A(t)$ denotes the ideal adiabatic evolution operator. This deviation function quantifies the instantaneous error between the actual and adiabatic trajectories throughout the evolution.

We rewrite the system Hamiltonian as
\begin{equation}
	H(t)=\frac{\Omega(t)}{2}(\ket{1(t)}\bra{1(t)}-\ket{2(t)}\bra{2(t)}),
\end{equation}
where $\ket{1(t)}$ and $\ket{2(t)}$ are the instantaneous eigenstates introduced previously.

To validate our theoretical framework, we consider three types of control paths on the parameter sphere:
\begin{align}
	&(\text{i}) ~~\theta(t)=\theta(0)+2c_\text{i}t,\varphi(t)=\varphi(0);\\
	&(\text{ii})~ \theta(t)=\theta(0),\varphi(t)=\varphi(0)+\frac{2c_\text{r}t}{\sin\theta};\\
	&(\text{iii})~ \theta(t)=\theta(0)+2c_\text{i}t,\varphi(t)=\varphi(0)+\int_{0}^{t}\frac{2c_\text{r}}{\sin\theta(t')}dt'.
\end{align}
These trajectories correspond to the conditions for constant nonadiabatic coupling $g_{1,2}(t) = C = c_\text{r} + i c_\text{i}$, with $g_{1,2}(t) = \left(\dot{\varphi}\sin\theta + i\dot{\theta}\right)/2$.

For a total evolution duration $\tau$, we define a set of time instants
\begin{equation}
t_j = \frac{(2j - 1)\tau}{2K}, \quad j = 1,2,\dots,K,
\end{equation}
where we choose $K = 5$ in our simulations. These $t_j$ mark the time points where phase kicks are introduced to ensure proper modulation of $\phi_{1,2}(t)$.

The remaining control parameter $\Omega(t)$ is engineered to enforce the phase condition
\begin{equation}
	\phi_{1,2}(t) = \int_0^t \left[\Omega(t') - \cos\theta(t')\,\dot{\varphi}(t')\right] \mathrm{d}t',
\end{equation}
which is modulated to flip between $0$ and $\pi$ (modulo $2\pi$) at the $t_j$ points. Specifically, we set
\begin{equation}
	\Omega(t) = \cos\theta(t)\,\dot{\varphi}(t), \quad \text{for } t \notin \cup_j [t_j - \delta_j, t_j],
\end{equation}
so that $\phi_{1,2}(t)$ remains constant. Within the short intervals $[t_j - \delta_j, t_j]$, we require a phase jump:
\begin{equation}
	\int_{t_j - \delta_j}^{t_j} \left[\Omega(t) - \cos\theta(t)\dot{\varphi}(t)\right] dt = \pi.
\end{equation}
Ideally, if $\Omega(t)$ could be made arbitrarily large, then the width $\delta_j$ could be made arbitrarily small. In practical settings, however, $\Omega(t)$ is bounded due to hardware constraints, and an appropriate $\delta_j$ must be selected.

In our simulations, we consider two scenarios:
\begin{itemize}
	\item \textbf{Ideal case:} We set $\Omega(t) = 2\pi\times25$ MHz during $\delta_j$ intervals, and choose $\delta_j = 0.02\,\mu$s.
	\item \textbf{Practical case:} We limit $\Omega(t)$ to approximately $2\pi\times1$ MHz, and set $\delta_j = 0.5\,\mu$s.
\end{itemize}
These parameter settings are compatible with current experimental capabilities in nitrogen-vacancy (NV) center platforms \cite{zhang2023coupling}.

\begin{figure}[!tbh]
	\centering
	\includegraphics[width=0.48\textwidth]{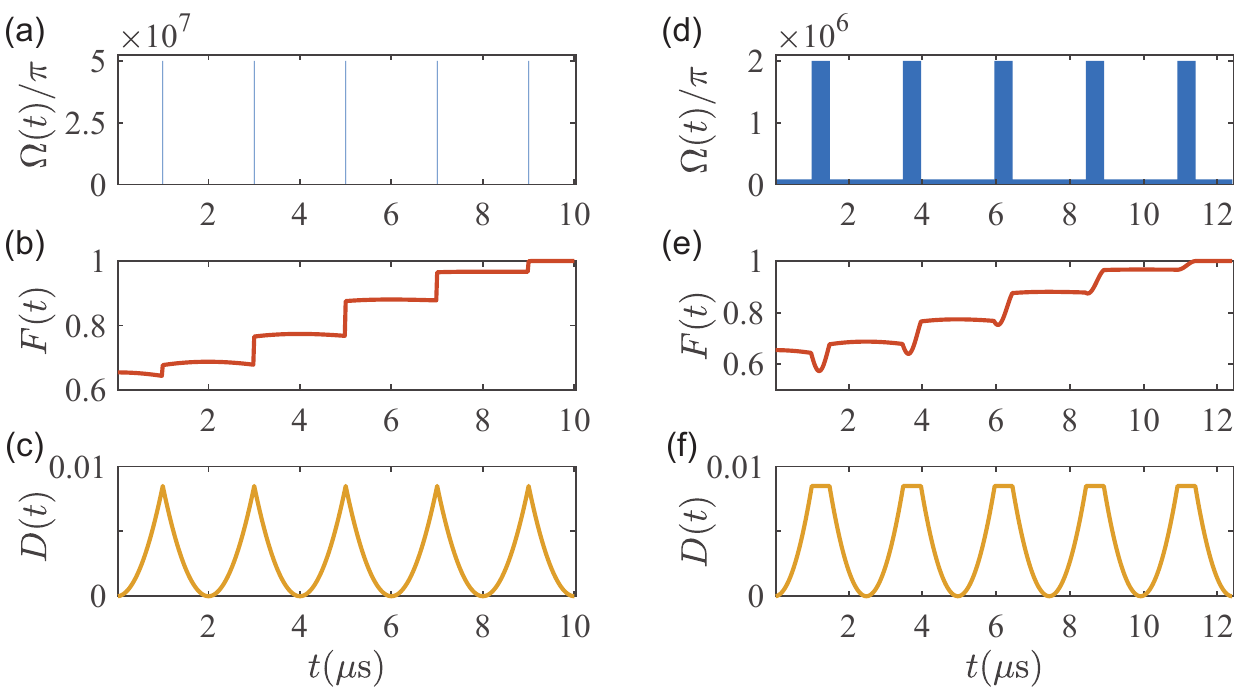}
	\caption{Acceleration of adiabatic evolution along the path in Case (ii). Panels (a)–(c) show the pulse amplitude, state fidelity, and adiabatic path deviation function for the ideal pulse amplitude scenario. Panels (d)–(f) present the corresponding results for the practical pulse amplitude scenario.}\label{fig:case2}
\end{figure}

For Case (i): We choose $\theta(0)=0$, $\theta(\tau)=\pi$, and a linear interpolation $\theta(t) = 2c_\text{i}t$, with $\varphi(t) = 0$ held constant. The initial state is $\ket{g}$, and the target final state is $\ket{e}$.

In the ideal scenario, the total evolution time is set to $\tau = 10.0~\mu$s. In the practical case, where the phase kick strength is limited to $\Omega = 2\pi\times1$ MHz and the width of the pulse is $\delta_j = 0.5~\mu$s, the total evolution time becomes $\tau' = 12.4~\mu$s.
In this case, all diagonal elements $g_{n,n}$ vanish, so the only applied controls are the phase-flipping Hamiltonians at the discrete time points $t_j$, which modulate $\phi_{n,m}$ in the intervals $[t_j - \delta_j, t_j]$ ($j = 1,2,\dots,5$). 

As shown in Fig.~\ref{fig:case1}, the state fidelity $F(t)$ reaches unity at the end of the evolution in both the ideal and practical cases. The adiabatic deviation function $D(t)$ in the ideal case exhibits sharp, triangular spikes, whereas in the practical case it shows plateau-like features with flattened peaks. By comparing Fig.~\ref{fig:case1}(d) and (f), one observes that the flat-topped structures arise from the finite-width $\pi$ pulses applied during $\phi_{n,m}$ flipping.

For Case (ii):  $\theta(t) = \pi/5$ remains constant, while $\varphi(t)$ evolves linearly from $0$ to $\pi$, i.e., $\varphi(t) = 2c_\text{r}t/\sin\theta$. The initial state is $\cos(\pi/10)\ket{g} + \sin(\pi/10)\ket{e}$, and the target state is $\cos(\pi/10)\ket{g} - \sin(\pi/10)\ket{e}$.

The total evolution time is $\tau = 10.0~\mu$s in the ideal case, and $\tau' = 12.4~\mu$s in the practical case. Here, the control field strength is slightly larger, with $\Omega \approx 2\pi\times1.0016$ MHz, due to the contribution from $\cos\theta\dot{\varphi} \approx 2\pi\times0.04$ MHz.
Unlike Case (i), the control Hamiltonian includes a continuous background component arising from $\cos\theta\dot{\varphi}$ in addition to the discrete $\pi$ pulses. This is clearly visible in Fig.~\ref{fig:case2}(d), where $\Omega(t)$ hovers around $2\pi\times1$ MHz. However, in the ideal case shown in Fig.~\ref{fig:case2}(a), this component is negligible compared to the dominant $2\pi\times25$ MHz pulse amplitudes.

For Case (iii): Both $\theta(t)$ and $\varphi(t)$ vary with time. Specifically, we choose $\theta(0) = \pi/5$, $\theta(\tau) = 4\pi/5$, with $\theta(t) = \pi/5 + 2c_\text{i}t$, and $\varphi(0) = 0$, $\varphi(\tau) = \pi$, with $\varphi(t) = \int_0^t 2c_\text{r}/\sin\theta(t')\,dt'$.

The initial state is $\cos(\pi/10)\ket{g} + \sin(\pi/10)\ket{e}$, and the target state is $\cos(2\pi/5)\ket{g} - \sin(2\pi/5)\ket{e}$. The total evolution time is $\tau = 10.0~\mu$s in the ideal case, and $\tau' = 12.4~\mu$s in the practical implementation with $\Omega = 2\pi\times1$ MHz and $\delta_j = 0.5~\mu$s.
In this case, since $\cos\theta(t)\dot{\varphi}(t) \approx 0$ throughout the evolution, the control Hamiltonian vanishes between the $\pi$ pulses. Consequently, the dynamics are entirely governed by the discrete $\phi_{n,m}$-flipping operations at times $t_j$.

\begin{figure}[!tbh]
	\centering
	\includegraphics[width=0.48\textwidth]{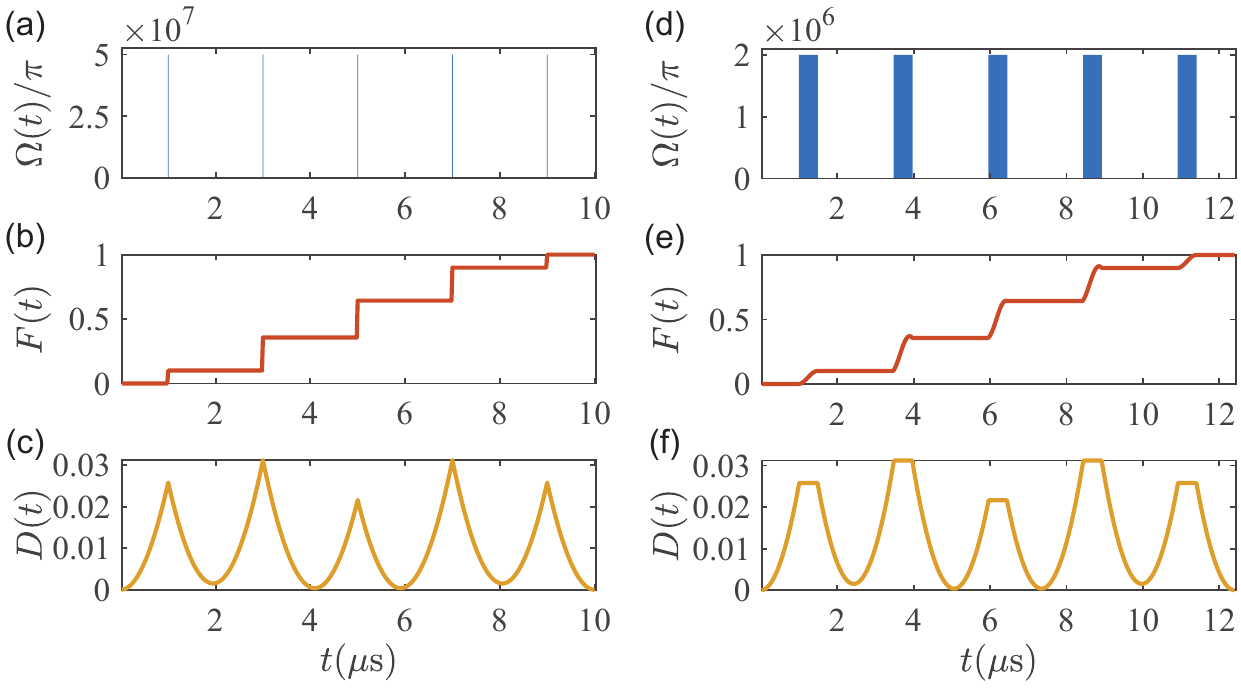}
	\caption{Acceleration of adiabatic evolution along the path in Case (iii). Panels (a)–(c) show the pulse amplitude, state fidelity, and adiabatic path deviation function under the ideal pulse condition. Panels (d)–(f) present the corresponding results for the practical pulse condition.}\label{fig:case3}
\end{figure}

\subsection{Three-level system}
\begin{figure}[!tbh]
	\centering
	\includegraphics[width=0.48\textwidth]{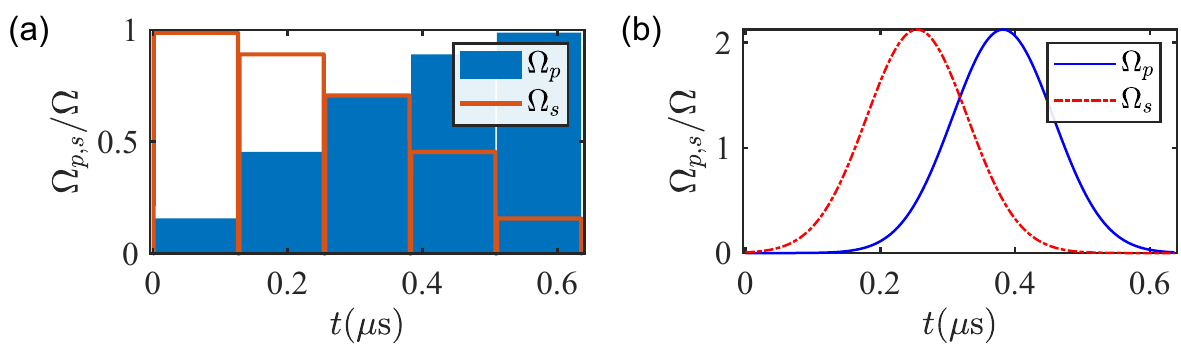}
	\caption{Pulse envelopes for the $\pi$-pulse and STIRAP schemes. (a) Pulse profiles of $\Omega_p$ and $\Omega_s$ corresponding to a sequence of five $\pi$ pulses. (b) Pulse profiles of $\Omega_p$ and $\Omega_s$ for the STIRAP, implemented over the same total duration as in (a) and with the same total pulse area.}\label{fig:pulse}
\end{figure}

We consider a three-level Hamiltonian
\begin{equation}
	H(t)=\frac{1}{2}\begin{pmatrix}
		0 & \Omega_p(t) & 0 \\
		\Omega_p(t) & 2\Delta & \Omega_s(t) \\
		0 & \Omega_s(t) & 0
	\end{pmatrix},
\end{equation}
written in the ordered basis $\{\ket{g},\ket{e},\ket{a}\}$.
The eigenstates of $H(t)$ can be expressed as 
\begin{align}
&\ket{1(t)}=\sin\theta\cos\varphi\ket{g}-\sin\varphi\ket{e}+\cos\theta\cos\varphi\ket{a}, \nonumber \\
&\ket{2(t)}=\cos\theta\ket{g}-\sin\theta\ket{a}, \nonumber \\
&\ket{3(t)}=\sin\theta\sin\varphi\ket{g}+\cos\varphi\ket{e}+\cos\theta\sin\varphi\ket{a},
\end{align}
with corresponding eigenvalues $E_1=-\frac{1}{2}\Omega(t)\tan\varphi$, $E_2=0$, and $E_3=\frac{1}{2}\Omega(t)\cot\varphi$. Here, we define the control amplitudes as $\Omega_p=\Omega(t)\sin\theta$, and $\Omega_s=\Omega(t)\cos\theta$, and use the relation $\tan2\varphi=\Omega(t)/\Delta$.

The nonadiabatic coupling terms $g_{n,m}$ are given by
\begin{align}
	& g_{1,1} = g_{2,2} = g_{3,3} = 0, \nonumber \\
	& g_{1,2} = -i\dot{\theta}\cos\varphi, \quad g_{1,3} = i\dot{\varphi}, \quad g_{2,3} = i\dot{\theta}\sin\varphi.
\end{align}
The corresponding $\phi_{n,m}$ are
\begin{align}
	& \phi_{1,2}(t) = -\phi_{2,1}(t) = \int_0^t -\frac{1}{2}\Omega(t')\tan\varphi\,dt', \nonumber \\
	& \phi_{2,3}(t) = -\phi_{3,2}(t) = \int_0^t -\frac{1}{2}\Omega(t')\cot\varphi\,dt', \nonumber \\
	& \phi_{1,3}(t) = -\phi_{3,1}(t) = \int_0^t -\frac{\Omega(t')}{\sin(2\varphi)}\,dt'.
\end{align}

To realize population transfer from $\ket{g}$ to $\ket{a}$, we set the initial parameters of the Hamiltonian as $\theta(0) = 0$ and $\varphi(0) = \pi/4$, so that the initial state $\ket{g}$ is the dark state $\ket{2(0)}$ of $H(0)$. By slowly varying $\theta(t)$ from $0$ to $\pi/2$ while keeping $\varphi$ fixed, the dark state adiabatically transforms from $\ket{g}$ to $\ket{a}$.
This adiabatic passage can be significantly accelerated using our pulse-based scheme, where $\pi$ pulses are applied along the evolution path. In our scheme, we apply different numbers of $\pi$ pulses ($K = 1$ to $5$) along the trajectory to suppress nonadiabatic transitions.

An alternative and widely-used method for efficient population transfer is the stimulated Raman adiabatic passage (STIRAP), which employs two time-dependent Gaussian pulses $\Omega_p(t)$ and $\Omega_s(t)$ to drive the system.
As we will show in the simulations, the success of the STIRAP protocol critically depends on how slowly the angle $\theta(t)$ is varied: longer evolution times yield higher final populations in state $\ket{a}$. To highlight the advantage of our scheme, we will compare its performance against STIRAP under equal total evolution times and pulse areas.

\subsection{Numerical simulations for three-level system}
\begin{figure*}[!tbh]
	\centering
	\includegraphics[width=1.0\textwidth]{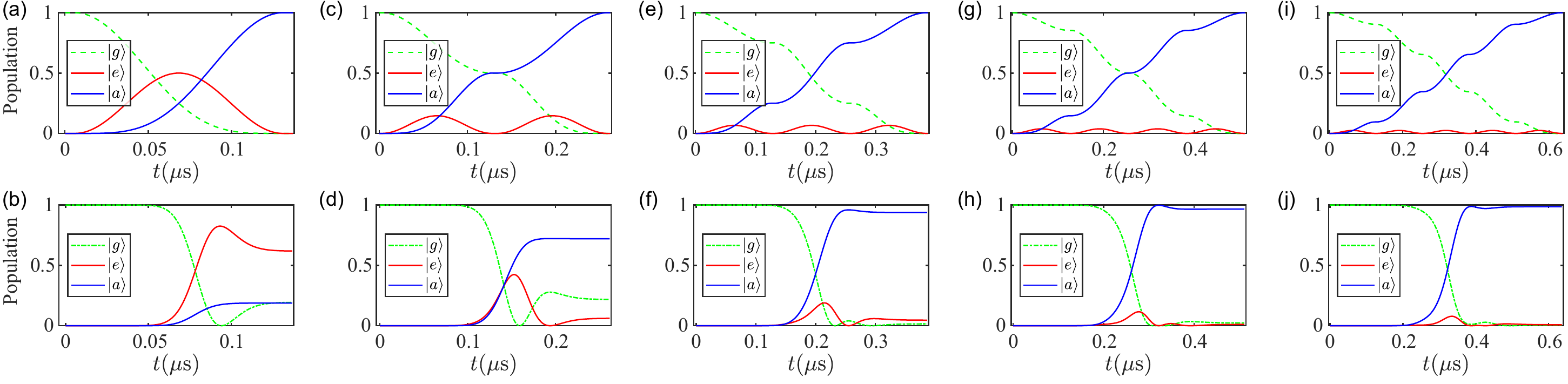}
	\caption{Population dynamics of the three energy levels under the $\pi$-pulse and STIRAP schemes. In all simulations, the initial state $\ket{g}$ is adiabatically transferred to the target state $\ket{a}$. The top row displays the population evolution for the $\pi$-pulse scheme with the number of applied $\pi$ pulses ranging from 1 to 5. Each subfigure in the bottom row presents the corresponding STIRAP, implemented with the same total duration and pulse area as its counterpart in the top row. The final population of $\ket{a}$ in (j) is 98.64$\%$.}\label{fig:3level}
\end{figure*}

A distinct difference between our scheme and STIRAP is the employed pulse envelops. As shown in Fig.~\ref{fig:pulse}, the applied pulses $\Omega_p$ and $\Omega_s$ in our scheme mimic the change in traditional adiabatic process (taking $N=5$ as an example), just replacing the continuous change used in traditional adiabatic process with the piece-wise change used in our scheme. On the other hand, the pulses used in the STIRAP are the Gaussian pulse.
For a fair comparison, the two schemes are implemented over the same time duration and with the same shape envelop area.

A key distinction between our scheme and the STIRAP protocol lies in the form of the applied pulse envelopes. As illustrated in Fig.~\ref{fig:pulse}, our scheme utilizes piecewise-constant pulses for $\Omega_p$ and $\Omega_s$. Taking $K=5$ as an example, these pulses approximate the continuous evolution path in the conventional adiabatic method by discretizing it into $K$ segments. In contrast, STIRAP employs smooth Gaussian pulses to achieve population transfer.
To ensure a fair comparison, both schemes are implemented over the same total evolution time and are normalized to have the same integrated pulse area.

In our scheme, $\pi$ pulses are applied at instants when the angle $\theta(t)$ satisfies $\theta(t) = (2j - 1)\pi/4K$ with $j = 1, 2, \ldots, K$. The pulse amplitude $\Omega$ is set to $2\pi \times 8$ MHz. Consequently, the total pulse area for both $\Omega_p(t)$ and $\Omega_s(t)$ in our scheme is given by
\begin{equation}
	\int_0^\tau \Omega_p(t)\,\mathrm{d}t = \int_0^\tau \Omega_s(t)\,\mathrm{d}t = 2\pi \times \sum_{j=1}^{K} \sin\theta_j.
\end{equation}

For the STIRAP scheme, the control pulses are defined as
\begin{align}
	\Omega_p^{S}(t) &= \Omega^{S} \exp\left[-\frac{(t - \tau/2 - \delta)^2}{\sigma^2}\right], \nonumber \\
	\Omega_s^{S}(t) &= \Omega^{S} \exp\left[-\frac{(t - \tau/2 + \delta)^2}{\sigma^2}\right],
\end{align}
where $\tau$ is the total evolution duration, $\delta = \tau/10$ is the relative delay between the pulses, and the standard deviation is set to $\sigma = \tau/6$. The peak amplitude $\Omega^S$ is chosen such that the integrated pulse areas match those in our scheme:
\begin{equation}
	\int_0^\tau \Omega_p^S(t)\,\mathrm{d}t = \int_0^\tau \Omega_s^S(t)\,\mathrm{d}t = 2\pi \times \sum_{j=1}^{K} \sin\theta_j.
\end{equation}

To evaluate the performance of the two schemes, we analyze the population dynamics of the three energy levels $\ket{g}$, $\ket{e}$, and $\ket{a}$ under varying evolution durations. As shown in Fig.~\ref{fig:3level}, the final population in state $\ket{a}$ achieved by STIRAP increases as the total evolution time becomes longer. For instance, when $\tau = 0.64\,\mu\text{s}$, the final population of $\ket{a}$ reaches $98.64\%$ (see Fig.~\ref{fig:3level}(j)). However, at a shorter duration of $\tau = 0.14\,\mu\text{s}$, which is already relatively long for practical quantum systems, the corresponding population is only $18.77\%$.

In contrast, our scheme achieves perfect population transfer to state $\ket{a}$ in all five cases shown in the top row of Fig.~\ref{fig:3level}, with the final population reaching $100.00\%$ regardless of $\tau$.
These results clearly demonstrate the superior efficiency of our pulse-based scheme over STIRAP. Furthermore, the maximum transient population in the intermediate excited state $\ket{e}$ is significantly lower in our scheme, indicating enhanced robustness to decoherence and other operational errors.

\section{Conclusion}
We would like to mention that during the preparation of this work, we became aware of a preprint \cite{liu2022shortcuts} reporting a result similar to our main conclusion. However, our work was conducted independently and prior to knowledge of this preprint. Importantly, our work provides a more detailed and rigorous derivation, offering new insights and a systematic framework that are not present in the preprint. We believe these aspects significantly enhance the understanding of the underlying mechanism and contribute substantial added value to the field.

In conclusion, we have proposed a general scheme to accelerate quantum adiabatic evolution by dynamically suppressing nonadiabatic transitions via $\pi$-pulse control. By decomposing the system's evolution operator into adiabatic and nonadiabatic components, we identified the condition under which nonadiabatic transitions can be effectively eliminated through piecewise constant control and $\pi$-pulse modulation. This method not only retains the inherent robustness of adiabatic processes but also significantly reduces the evolution time.

We demonstrated the feasibility and performance of our scheme in both two-level and three-level quantum systems. Compared to the widely used STIRAP protocol, our scheme yields higher target-state fidelity and better suppression of nonadiabatic transitions within the same evolution time. Furthermore, the scheme is well-suited for systems with energy-level crossings or minimal energy gaps, and it provides the flexibility to bypass unwanted segments along the evolution path.

Our work provides a new perspective for achieving rapid and robust quantum control and can be readily extended to more complex quantum systems. Future research may explore its applications in high-dimensional Hilbert spaces, quantum state engineering, and scalable quantum information processing platforms.

\begin{acknowledgments}
	T.X. acknowledges support by the National Natural Science Foundation of China (No. 12247165).
	J.Z. acknowledges support by the National Natural Science Foundation of China (No. 12004206).
	G.L. acknowledges support by the National Natural Science Foundation of China (No. 62471046), Beijing Advanced Innovation Center for Future Chip (ICFC), and Tsinghua University Initiative Scientific Research Program.
\end{acknowledgments}

\bibliography{jumpreference}

\end{document}